\documentstyle[a4,11pt]{article} 

\newcommand{\sect}[1]{\setcounter{equation}{0}\section{#1}}
\newcommand{\subsect}[1]{\subsection{#1}}

\newcommand{\vs}[1]{\rule[- #1 mm]{0mm}{#1 mm}}
\newcommand{\hs}[1]{\hspace{#1 mm}}

\newcommand{\lbl}[1]{\label{eq:#1}}
\newcommand{\rf}[1]{(\ref{eq:#1})}
\newcommand{\nn}{\nonumber}

\newcommand{\be}{\vs{2}\large\begin{equation}}
\newcommand{\ee}{\vs{2}\end{equation}\normalsize}
\newcommand{\cee}{\vs{2}\]\normalsize}
\newcommand{\cbe}{\vs{2}\large\[}
\newcommand{\bea}{\large\begin{eqnarray}}
\newcommand{\ena}{\end{eqnarray}\normalsize}
\newcommand{\nnbea}{\large\begin{eqnarray*}}
\newcommand{\nnena}{\end{eqnarray*}\normalsize}

\def\ie{{\em i.e. }}
\def\eg{{\em e.g. }}
\def\etal{{\em et al. }}

\newcommand{\lp}{\left(}
\newcommand{\rp}{\right)}
\newcommand{\blp}{\biggl(}
\newcommand{\brp}{\biggr)}

\newcommand{\lra}{\ \longrightarrow\ }
\newcommand{\Lra}{\ \Longrightarrow\ }
\newcommand{\Llra}{\ \Longleftrightarrow\ }
\newcommand{\ovl}[1]{\overline{#1}}

\newcommand{\pt}{\! \cdot \!}

\newcommand{\sm}[2]{\textstyle{\frac{#1}{#2}}\displaystyle}
\newcommand{\shalf}{\textstyle{\frac{1}{2}}\displaystyle}
\newcommand{\dps}{\displaystyle}

\newcommand{\hcm}{\hspace{1cm}}

\newcommand{\br}[1]{{\ovl{#1}}}
\newcommand{\bz}{{\ovl{z}}}
\newcommand{\bZ}{\br{Z}}
\newcommand{\zbz}{(z,\bz)}
\newcommand{\ZBZ}{(Z,\ovl{Z})}
\newcommand{\cD}{{\cal D}}
\newcommand{\cR}{{\cal R}}
\newcommand{\mmb}{\mu\,\ovl{\mu}}
\newcommand{\paz}{\partial_z}
\newcommand{\paZ}{\partial\hs{-1}\vs{1.5}_Z}
\newcommand{\prt}{\partial} 
\newcommand{\prtb}{\br{\partial}}
\newcommand{\pab}{\partial_{\,\bz}}
\newcommand{\Pab}{\partial\vs{1.5}_\bZ}

\newcommand{\go}{\stackrel{\circ}{g\,}}

\newcommand{\Fo}{\stackrel{\circ}{\phi\,\,}\!\!\!}
\newcommand{\ro}{\stackrel{\circ}{\rho\,}}
\newcommand{\Ro}{\!\stackrel{\circ}{R\,\,}\!\!}
\newcommand{\rZ}{\rho\hs{-1.2}\vs{2}_{Z\bZ}}
\newcommand{\roZ}{\stackrel{\circ}{\rho\,}\hs{-2.5}\vs{2}_{Z\bZ}}

\newcommand{\mub}{\br{\mu}}
\newcommand{\anom}{\mbox{\huge {\em a}}}
\newcommand{\cdotd}{(c\pt\prt)}

\newcommand{\GAM}{\mbox{\Large $\displaystyle \Gamma$}}
\newcommand{\mes}[2]{{\mbox{\small $\displaystyle \frac{d \br{#1} \wedge d#2}
{2i}$}} \ }
\newcommand{\zer}[1]{\stackrel{\circ}{#1}}

\newcommand{\NPB}[1]{Nucl.\ Phys.\ {\bf B#1}}

\newcommand{\PLB}[1]{Phys.\ Lett.\ {\bf B#1}}

\newcommand{\CMP}[1]{Commun.\ Math.\ Phys.\ {\bf #1}}

\newcommand{\LMP}[1]{Lett.\ Math.\ Phys.\ {\bf #1}}
\newcommand{\PR}[1]{Phys.\ Rev.\ {\bf #1}}

\newcommand{\MPL}[1]{Mod.\ Phys.\ Lett.\ {\bf #1}}
\newcommand{\IJMP}[1]{Inter. Journ. Mod.\ Phys.\ {\bf #1}}

\begin{document}

\font\fifteen=cmbx10 at 15pt
\font\twelve=cmbx10 at 12pt

\begin{titlepage}

\begin{center}

\renewcommand{\thefootnote}{\fnsymbol{footnote}}

{\twelve Centre de Physique Th\'eorique\footnote{
Unit\'e Propre de Recherche 7061
}, CNRS Luminy, Case 907}

{\twelve F-13288 Marseille -- Cedex 9}

\vskip 3.cm

{\fifteen Integrating the Chirally Split Diffeomorphism Anomaly\\[2mm]
on a Compact Riemann Surface}
\footnote{Partially based on a talk 
``{\em Holomorphic Projective Connections and the
 Liouville Equation}" presented at the 1-st worshop of Mathematical
 Physics, on ``Quantization of Liouville Theory and Related Models",
 12-17 January 1998, Montpellier, France.}

\vskip 1.5cm

\setcounter{footnote}{0}
\renewcommand{\thefootnote}{\arabic{footnote}}

{\bf 
Serge LAZZARINI
\footnote{and also Universit\'e de la M\'editerran\'ee, Aix-Marseille
II, e-mail : {\tt sel@cpt.univ-mrs.fr}}
}
\end{center}

\vskip 2.cm

\centerline{\bf Abstract}

A well-defined chirally split functional integrating the 2D
chirally split diffeomorphism anomaly is exhibited on an arbitrary compact
Riemann surface without boundary. The construction requires both
the use of the Beltrami parametrisation of complex structures and
the introduction of a background metric possibly subject to a
Liouville equation. This formula reproduces in the flat case the
so-called Polyakov action. Althrough it works on the torus (genus 1),
the proposed functional still remains to be related to a 
Wess-Zumino action for diffeomorphisms.

\vskip 1.5cm

\noindent 1995 PACS Classification: 04.60.K, 11.25.Hf, 11.10.Gh

\indent

\noindent Key-Words: Polyakov action, Beltrami
parametrisation of complex structures, Liouville theory.

\indent

\noindent {\today}

\noindent {CPT-98/P.3637, to be published in Phys. Lett. B}

\indent

\noindent anonymous ftp or gopher: cpt.univ-mrs.fr

\end{titlepage}
\setcounter{footnote}{0}

\sect{Introduction}

\indent

In the last decade various attempts have been made in order to find out a
chirally split Polyakov action \cite{Pol87} emerging from the study on
the quantization of 2-dimensional gravity
which is conformally covariant on a compact Riemann surface without boundary
\cite{Yos89,BBG89,Laz90,Gri90,Hab90,Yos90,Zuc91,Zuc93a,AT97}.

Since another celebrated paper by Polyakov \cite{Pol81} some partial
results have been disseminated throughout several relevant papers
\footnote{The author apologizes to all those who have been only
implicitly referred to, in the bibliography.}. 
Quite recently there has been a revival interest in the subject of
Liouville theory fifteen years later and in the relationship with 
the chirally split 
Polyakov effective action for induced 2-d gravity \cite{AT97} as well.
As shown in \cite{KLS91a,KLS91b}, the existence of a class of such functionals
has been proved by using the local index theorem for families of
elliptic operators in the form given by Bismut \etal \cite{BGS88}, and
as well as their supersymmetric version \cite{Gie93b}. It is
emphasized that the author's view point falls into the perturbative
approach to the chiral splitting property of bidimensional conformal
models at {\em non} vanishing central charge as a local strengthened
form of the Belavin-Knizhnik holomorphic factorization theorem
\cite{BK86a,BK86b}. 

In this Letter a globally defined `Polyakov formula' is proposed
althrough it is naively derived from the Liouville action proposed in
\cite{Pol81}. In doing so by using the Beltrami parametrization of
complex structures, it will be found a well-defined holomorphically factorized
functional integrating the globally defined factorized anomaly for
diffeomorphims on an arbitrary compact Riemann surface without boundary.
The related energy-momentum tensor will be discussed.

\sect{The smooth change of complex coordinates}

\indent

Throughout the paper we will be concerned with the euclidean framework.
We shall right away introduce a complex analytic atlas 
with local coordinates $\zbz$ 
\footnote{We shall omit the index denoting the open set where these
coordinates are locally defined since all formulae will glue through 
holomorphic changes of coordinates.}
corresponding to the reference complex structure defined by a
background metric $\go$ \cite{BBS86,Bec88}, namely one has the following 
equality of quadratic forms,
\be
d\!\zer{s}{}^2\ =\ 
\go_{\hs{-1.2}\alpha \beta} \, dx^{\alpha}dx^{\beta}\equiv\
\ro_{\hs{-1.2}z\bz} \,|dz|^2\ ,
\lbl{dso}
\ee
where $\ro_{\!z\bz}\equiv\ro>0$ is the coefficient of a positive 
real valued type (1,1) conformal background field. We then parametrize
locally the metric $g$ according to \cite{Leh87},
\be
ds^2 \ =\ g_{\alpha \beta} \, dx^{\alpha}dx^{\beta} \ =\ \rho_{z\bz} \,
|dz + {\mu_{\bz}}^z d\bz|^2 \ ,
\lbl{dsz}
\ee
where ${\mu_{\bz}}^z\equiv\mu$ 
is the local representative of the Beltrami differential 
$(|\mu|<1)$ seen as a $(-1,1)$ conformal field, which parametrizes
the conformal class of the metric $g$ and $\rho_{z\bz}\equiv\rho>0$ 
is the coefficient of a positive real valued type (1,1) conformal field. 

Our compact Riemann surface $\Sigma$ without boundary being now 
endowed with the analytic atlas with local coordinates $\zbz$, 
let $(Z,\bZ)$ be the local coordinates of another holomorphic atlas
corresponding to the Beltrami differential $\mu$,
\be
dZ \ =\ \lambda (dz + \mu d\bz) \hcm \Lra
\Pab \ =\ \frac{\prtb - \mu \prt}{\br{\lambda}(1 - \mmb)}
\lbl{dZ}
\ee
where $\lambda\ =\ \prt_z Z\ \stackrel{\mbox{\tiny Def}}{\equiv}\,\prt Z$
is an integrating factor fulfilling,\footnote{From now on, 
we shall reserve $\prt$'s for $\prt\equiv\paz,\ \prtb\equiv\pab$.}
\be
(d^2\ =\ 0) \Llra 
(\prtb - \mu \prt) \ln \lambda \ = \ \prt \mu \ .
\lbl{lam}
\ee
Solving the above Pfaff system \rf{dZ} is equivalent to solving locally 
the so-called Beltrami equation,
\be
(\prtb - \mu\prt)\, Z\ =\ 0\ .
\lbl{beltra}
\ee
According to Bers, see \eg \cite{Leh87}, the Beltrami equation
\rf{beltra} always admits as a solution a quasiconformal mapping
with dilatation coefficient $\mu$. One thus remarks that $Z$
is a (non-local) holomorphic functional of $\mu$ as well as the
integrating factor $\lambda$, and will be seen to serve as a
parametrization of a family of global diffeomorphisms connected to 
the identity \cite{Sto88que} in analogy to what it is thought about
the ``light-cone" gauge together with the ``chiral diffeomorphisms"
\cite{Pol87,GN89}. However, the solution of the Beltrami equation 
define a smooth diffeomorphism $ \zbz \lra \ZBZ$ which 
preserves the orientation (the latter condition secures the 
requirement $|\mu|<1$, \eg \cite{Zuc93a}), so that 
$\ZBZ$ defines a new system of 
complex coordinates with $Z\lra z$ when $|\mu|\lra 0$.

In terms of the $\ZBZ$ complex coordinates which by virtue of \rf{dZ}
turns out to be isothermal coordinates for the metric $g$ by defining
the non-local metric,
\be
\rZ \equiv \frac{\rho}{\lambda\,\ovl{\lambda}}\ .
\lbl{rZ}
\ee
In particular, the quadratic form \rf{dsz}, the volume form, 
the scalar curvature and the scalar Laplacian, respectively write,
\bea
ds^2 \ =\ g_{\alpha \beta} \, dx^{\alpha}dx^{\beta}\ =\ \rZ\,|dZ|^2,
&& \sqrt{g}\ =\ \rho\,(1-\mmb),\nn\\
&&{}\lbl{dsZ}\\[-2mm]
d^2\!x \sqrt{g} = \mes{Z}{Z}\rZ,\hcm
R(g) = \frac{-4}{\rZ}\,\paZ\Pab \ln\!\rZ,&& \Delta(g)= 
\frac{-4}{\rZ}\paZ\Pab\ .\nn
\ena

\sect{The locally defined Weyl$\times$Diff BRS algebra and the WZ
consistency conditions}

\indent

The corresponding locally defined BRS algebra 
is by now well-known \cite{BB87,Sto87,Bec88,KLT90},
\be\begin{array}{l}
s \ln\rho\ =\ \Omega + \nabla c + \br{\nabla}\br{c} + \mu\,\prt\br{c}
+ \mub\,\prtb c\ ,\hcm s\!\ro\ =\ 0\ ,\\[2mm]
s \Omega\ =\ \cdotd \Omega\ , 
\hcm s c\ =\ \cdotd c\ ,\hcm \mbox{and c.c.}\\[2mm]
s \mu\ =\ \prtb C + C \prt \mu - \mu \prt C\ ,\hcm \mbox{and c.c.} \\[2mm]
s\ln\lambda\ =\ \prt C + C\,\prt\ln\lambda \equiv \cD\,C\ ,
\hcm \mbox{and c.c.}\\[2mm]
s C\ =\ C \prt C \ ,\hcm \mbox{and c.c.}
\end{array}
\lbl{BRSz}
\ee 
where $C = c + \mu \br{c}$ \cite{Bec88} is the combination of the
diffeomorphism ghost $(c,\br{c})$ appropriate to the complex analytic
structure of Diff$_0(\Sigma)$ (the connected component to the
identity) and where,
\be
\nabla c \ =\ \prt c + c\,\prt\ln\!\rho \ ,
\lbl{del}
\ee
is the covariant derivative related to the Weyl factor $\rho$ of the
metric $g$. Of course $s^2=0$.

One may extend the $s$-operation in the $\ZBZ$ of complex coordinates system
by, see \eg \cite{Laz97},
\be
sZ = c^Z \ \equiv \ \lambda C, \hcm
s\lambda = \prt c^Z, \hs{5} \& \ \mbox{c.c.} \hcm
s\ln\rZ = \Omega + (c^Z\paZ+c^\bZ\Pab)\ln\rZ\ ,
\lbl{sZ}
\ee
which will be used troughout to simplify some computational steps.

\indent

On the one hand, the standard Weyl anomaly \cite{Pol81} 
reflecting the famous trace anomaly \cite{BC-RR83} writes
\be
\anom(\Omega,\rho,\mu,\mub)\ =\ 
- \sm{k}{12\pi}\int_{\Sigma} \mes{Z}{Z} \Omega\, \paZ\Pab \ln\!\rZ\ ,
\lbl{WaZ}
\ee
and on the other hand, the chirally split form of the diffeomorphism
anomaly reads \cite{LS89,KLT90}
\cbe
\anom(C,\mu;{\cal R}) + \br{\anom(C,\mu;{\cal R})}\ ,
\cee
with
\be
\anom(C,\mu;{\cal R})\ =\ \sm{k}{12 \pi} \int_{\Sigma} \mes{z}{z} 
C \lp \prt^3 + 2{\cal R}\prt + \prt{\cal R} \rp \mu \ ,
\lbl{anom}
\ee
where $\cal R$ is the representative of a background holomorphic
projective connection \cite{Gun66}, and the Wess-Zumino consistency 
conditions \cite{WZ71} for these two forms of the same anomaly 
phenomemon now reads,
\cbe
s\anom(\Omega,\rho,\mu,\mub)\ =\ s\anom(C,\mu;{\cal R})\ =\ 0.
\cee

\sect{Towards the holomorphic factorization of the gravitational action}

\indent

In the spirit of the genuine Wess-Zumino action, the smooth change of
coordinates \rf{dZ} might be seen as already said before like
a parametrisation of a
family of global diffeomorphisms connected to the identity \cite{Sto88que}. 
Afterwards, roughly speaking, one will argue since the effective action has to
break the diffeomorphism invariance, \ie passing from the
$\zbz$ system of complex coordinates to the $\ZBZ$ one is not
covariantly performed. 
The important feature will be that the effective action remains
well-defined in the $\zbz$ background coordinates.
Instead of using the zweiben formalism \cite{SSW89,BB87}, 
the covariant effective action for 2d gravity \cite{Pol81,Pol87}
\be
\Gamma[g] = \sm{k}{96\pi} \int_{\Sigma} d^2\!x 
\sqrt{g}\, R(g) (-\Delta(g))^{-1} R(g)\ ,
\ee
is promoted in the $\ZBZ$ coordinate system, by using formally 
the replacements given in \rf{dsZ}. One formally has
\be
\GAM[\mu,\mub,\rho] = - \sm{k}{12\pi}\int_{\Sigma} \mes{Z}{Z}
\shalf \paZ\Pab \ln\!\rZ\, \frac{1}{\paZ\Pab}\,\paZ\Pab \ln\!\rZ \ .
\lbl{boxZ}\\[2mm]
\ee
Of course, the Goldstone scalar field $\Phi$ `$=$' 
$(-\Delta(g))^{-1} R(g)$ is expected to be a
non-local functional in the metric $g$ and thus in the Beltrami
coefficient $\mu$. 
Firs of all, recall that on a compact Riemann surface
the operator $\paZ\Pab$ maps scalar fields to
conformal fields of weight (1,1) with
\be
\paZ\Pab\ : \Omega^{0,0}/\mbox{\bf R}\lra \Im
\equiv\{\omega\in\Omega^{1,1},\ \int_\Sigma \omega = 0\}.
\ee
the constants being the zero modes, and the image is so defined in
order to insure the inverse operator. It is noticed that $\paZ\Pab
\ln\!\rZ$ does not belong to the image of $\paZ\Pab$.
However, in order to have an idea of this non-local 
dependence let us be now very naive by writing
\be 
\frac{1}{\paZ\Pab}\,\paZ\Pab \ln\!\rZ\ \mbox{`$=$'}\ \ln\!\rZ\ ,
\lbl{bad}
\ee
such that the functional \rf{boxZ} turns out to be
 nothing but the usual Liouville action
with no cosmological term. The latter is manifestly not 
covariantly written it the $\ZBZ$ system of complex coordinates and
where the related Liouville field as introduced in \cite{Pol81} 
\be
\phi\ZBZ\ =\ \ln \rZ\ZBZ,\hcm
\lbl{PL}
\phi\ZBZ\ =\ \phi(W,\br{W}) + \ln\left|\frac{dW}{dZ}\right|^2,\hcm
W=W(Z)\ ,
\ee
is a non-local functional in the complex structure $\mu$. 
In some sense, in this very naive approach 
the ill-definiteness in the $\ZBZ$ complex
coordinates of the action \rf{boxZ} into which \rf{bad} has been plugged
does not matter beacause, fortunately, it will be conformally covariant 
in the background $\zbz$ system of complex coordinates.
On the one hand, one has that
\be
\phi\zbz\ =\ \blp\ln\!\rho-\ln\lambda - \ln\ovl{\lambda}\brp\zbz
\lbl{Liouphi}
\ee
is a well-defined chirally split non-local scalar field 
in the background complex structure with
\be
s\,\phi\zbz\ =\ \Omega\zbz + (\cdotd\Phi)\zbz\ ,
\ee
by virtue of eq.\rf{lam}. And on the other hand, formula
\rf{boxZ} now rewrites in a globally defined way on a
compact Riemann surface without boundary,
\be
\GAM[\mu,\mub,\rho] = \sm{k}{24\pi}\int_{\Sigma} \mes{z}{z}
\!\ln\!\lp\frac{\rho}{\lambda \ovl{\lambda}}\rp\!
\lp\!(\prt-\prtb\mub)\frac{\nabla\!\mu}{1-\mu\mub} + 
(\prtb-\prt\mu)\frac{\ovl{\nabla}\!\mub}{1-\mu\mub} -
\prt\prtb\ln\!\rho\rp\!,
\ee
after using the following decomposition for the symmetric scalar Laplacian 
\cite{Laz90},
\be
\mes{Z}{Z}\paZ\Pab = \mes{z}{z}\lp \prt\prtb - 
(\prt-\prtb\mub)(1-\mu\mub)^{-1} \mu \prt - 
(\prtb-\prt\mu)(1-\mu\mub)^{-1} \mub\, \prtb\rp,
\lbl{lapl}
\ee
and with the covariant derivative \rf{del}, 
$\nabla\mu=\prt\mu+\mu\prt\ln\!\rho$, acting on $\mu$. In the course of the
computation it is crucial to use eq.\rf{lam} for the integrating
factor. Since it is just a change of coordinates, one easily checks,
using the algebra \rf{BRSz} (or \rf{sZ} combined with \rf{lapl}), that
(up to globally defined surface terms),
\cbe
s\,\Gamma[\mu,\mub,\rho] \ =\ \anom(\Omega,\rho,\mu,\mub)
\cee
the Weyl anomaly \rf{WaZ} in the $\ZBZ$ complex coordinates.

\subsect{A well-defined chirally split action integrating the 
diffeomorphism anomaly on a compact Riemann surface}

\indent

Using the well-defined Liouville action proposed in \cite{KLT90},
\be
\GAM_{\!\mbox{\tiny Liouv}}(\varphi,g)\ =\ 
- \sm{k}{12\pi}\int_{\Sigma} \mes{Z}{Z}
\lp\shalf \varphi\paZ\Pab\,\varphi + \varphi \paZ\Pab \ln\!\rZ\rp \ ,
\lbl{LZ}\ee
where the new scalar Liouville field is \cite{KLT90}, 
\be
\varphi\equiv\ln\roZ - \ln\rZ=\ln\!\ro-\ln\rho, \hcm
\roZ\equiv \frac{\ro}{\lambda\ovl{\lambda}}\ .
\lbl{roZ}
\ee
The sum $\GAM[\mu,\mub,\rho]+\GAM_{\!\mbox{\tiny Liouv}}(\varphi,g)$
corresponding to an effective action for 2d gravity in a background, yields
(after dropping a well-defined surface term), 
\be
\GAM[\mu,\mub;\ro] = - \sm{k}{12\pi}\int_{\Sigma} \mes{Z}{Z}
\lp\shalf \ln\!\roZ\, \paZ\Pab \ln\!\roZ\rp ,
\lbl{box}\\[2mm]
\ee
for which the same comments made for \rf{boxZ} apply and where now
the related Liouville field depending in the background $\ro$,
\be
\Fo\ZBZ\ =\ \ln \roZ\ZBZ  \ ,
\lbl{PLo}
\ee
is a non-local functional in the complex structure $\mu$. 
We stress again that the ill-definiteness in the $\ZBZ$ complex
coordinates does not matter beacause,
fortunately, one can assert that \rf{box} is conformally covariant 
in the background $\zbz$ system of complex coordinates in which now,
on the one hand\footnote{K. Yoshida in \cite{Yos90} used $\ro=1$.},
\be
\Fo\zbz\ =\ \blp\ln\!\ro-\ln\lambda - \ln\ovl{\lambda}\brp\zbz
\lbl{Liouphio}
\ee
is a well-defined chirally split, but non-local, scalar field 
in the $\zbz$ complex structure, and on the other hand, as above, formula
\rf{box} now rewrites in a globally defined way on a
compact Riemann surface without boundary,
\be
\GAM[\mu,\mub;\ro] = \sm{k}{24\pi}\int_{\Sigma} \mes{z}{z}
\!\ln\!\lp\frac{\ro}{\lambda \ovl{\lambda}}\rp\!
\lp\!(\prt-\prtb\mub)\frac{\zer{\nabla}\!\mu}{1-\mu\mub} + 
(\prtb-\prt\mu)\frac{\zer{\ovl{\nabla}}\!\mub}{1-\mu\mub} -
\prt\prtb\ln\!\ro\rp\!,
\ee
after using once more \rf{lapl}, and where the covariant derivative 
$\zer{\nabla}\mu=\prt\mu+\mu\prt\ln\!\ro$ related to the
background metric has been introduced. As before
it is crucial to use eq.\rf{lam} for the integrating factor.
This functional may be considered as the coupling of the
``diffeomorphism-Goldstone scalar boson" $\Fo$ to the gauge fields
$\mu$ and $\mub$. One also checks that 
\be
s\,\GAM[\mu,\mub;\ro]\ =\
\anom(c,\bar{c},\mu,\mub;\ro)\equiv \frac{-k}{12\pi}\int_{\Sigma} 
\mes{Z}{Z} \Omega_{\mbox{\tiny Comp}} \paZ\Pab\ln\!\roZ\ ,
\ee
the non-chirally split diffeomorphism anomaly given in \cite{KLT90}
depending on the background metric $\ro$ and with the compensating ghost
${\dps \Omega_{\mbox{\tiny Comp}}\equiv - \blp 
\zer{\nabla}\! c + \zer{\br{\nabla}}\!\br{c} + \mu\,\prt\br{c} +
\mub\,\prtb c\brp}$.

Then performing well-defined integrations by parts and dropping terms
of the form d$\chi$, with $\chi$ a globally defined non-local one form
on the compact Riemann surface, it is straightforward to find
\bea
\GAM[\mu,\mub;\ro] &=& \sm{-k}{24\pi}\int_{\Sigma} \mes{z}{z}\!\lp 
\!\ln\!\lp\frac{\ro}{\lambda \ovl{\lambda}}\rp\!\prt\prtb\ln\!\ro
+(\prt\ln\!\ro-\prt\ln\lambda)\zer{\nabla}\!\mu 
+(\prtb\ln\!\ro-\prtb\ln\ovl{\lambda})\zer{\ovl{\nabla}}\!\mub \rp \nn\\
&&\hs{10}
+  \underbrace{
\sm{k}{12\pi}\int_{\Sigma} \mes{z}{z}(1-\mu\mub)^{-1}
\lp\zer{\nabla}\!\mu\zer{\ovl{\nabla}}\!\mub
-\shalf\mub(\zer{\nabla}\!\mu)^2
-\shalf\mu(\zer{\ovl{\nabla}}\!\mub)^2\rp}_{{\dps \equiv
-\GAM_{\!\mbox{\tiny III}}(\ro,\mu,\mub)}} .\lbl{fact}
\ena
The last line exactly provides the globally defined local counterterm
$\GAM_{\!\mbox{\tiny III}}(\mu,\mub;\ro)$ already found in
\cite{KLT90} which is the appropriate form of that proposed in
\cite{SSW89,VV89,Ver90}, insuring its global definition on an 
arbitrary compact Riemann surface without boundary. It is a necessary
counterterm in order to obtain a chirally split form of the
diffeomorphism anomaly. Once more
eq.\rf{lam} for the integrating factor has been used.
One can directly write out an holomorphically factorized effective
action with a background metric by modifying (up to well-defined
surface terms) the first integral in \rf{fact} as
\bea
&&\GAM_{\!\mbox{\tiny Chiral}}[\mu,\mub;\ro] \equiv \GAM[\mu,\mub;\ro] +
\GAM_{\!\mbox{\tiny III}}(\mu,\mub;\ro) \ =\ 
\sm{k}{12\pi}\int_{\Sigma} \mes{z}{z}\!\lp 
\!\shalf\ln\!\lp\frac{\ro}{\lambda \ovl{\lambda}}\rp\!\prt\prtb\ln\!\ro
\right. \nn\\[2mm]
&&\hs{-15} \left. +\ \mu(\Ro-\{Z,z\}) + \shalf
\prtb\ln\lambda\, (\prt\ln\!\ro - \prt\ln\lambda) 
+ \mub(\zer{\ovl{R}\,}-\{\bZ,\bz\}) + \shalf
\prt\ln\ovl{\lambda}\, (\prtb\ln\!\ro - \prtb\ln\ovl{\lambda})\rp,
\lbl{WZP}
\ena
where both the curvature, 
$\Ro\ =\prt^2\ln\!\ro - \shalf (\prt\ln\!\ro)^2$,
and the Schwarzian derivative of $Z$ with respect to $z$, 
$\{Z,z\}=\prt^2\ln\!\lambda - \shalf (\prt\ln\!\lambda)^2$, 
explicitly appear. Remark that due to the first term in the integrand
of \rf{WZP}, $\GAM_{\!\mbox{\tiny Chiral}}[\mu,\mub;\ro]$
can not be chirally split into the sum of two well-defined
functionals, namely,
\be
\GAM_{\!\mbox{\tiny Chiral}}[\mu,\mub;\ro]\ \neq\ 
\GAM_{\!\mbox{\tiny Chiral}}[\mu,0;\ro] + 
\GAM_{\!\mbox{\tiny Chiral}}[0,\mub;\ro]\ ,
\lbl{split}
\ee
whereas clearly 
$\delta_{\mu}\delta_{\mub}\GAM_{\!\mbox{\tiny Chiral}}[\mu,\mub;\ro] = 0$.
The former is related to the Liouville action for the field
$\phi$ as will be shown down below, (see formula \rf{WZP'}).

One has thus exhibited a class of Polyakov-like formulae 
as non-local functionals in the 
complex structure $\mu$ and depending on a background metric $\ro$. 
The usual formula in the flat case \cite{Pol87}
is recovered by setting $\ro=1$.

Furthermore, a direct comparison with the other proposal for a Polyakov
action given by R.~Zucchini shows that the integrating factor
$\lambda$ plays the role of the $\mu$-holomorphic $(1,0)$-differential
introduced in \cite{Zuc91,Zuc93a}. But it has the great advantage to
never vanish since it is half of the Jacobian of the smooth change of
complex variables $z\lra Z$. Moreover the holomorphic background 
$(1,0)$-differential is replaced by a {\em non-vanishing} background 
$(1,1)$-differential, namely the
background metric $\ro>0$. However the Polyakov action as proposed by
Zucchini can be written in a chirally split form (equals sign in \rf{split}).

Modulo a well-defined surface term,
one directly checks that the chirally split diffeomorphism anomaly
depending on the background metric $\ro$ is re-obtained by,
\be
s\,\GAM_{\!\mbox{\tiny Chiral}}[\mu,\mub;\ro]\ =\ \anom(C,\mu;\ro) \ 
+\ \overline{\anom(C,\mu;\ro)} \ ,
\lbl{a-plit0}
\ee
with
\be
\anom(C,\mu;\ro)\ =\ \sm{k}{12 \pi} \int_{\Sigma} \mes{z}{z} 
C \lp \prt^3 \mu - (\prtb - \mu \prt - 2\prt \mu ) \Ro \rp.
\lbl{a-0}
\ee
However note that this action integrating the chirally split
diffeomorphism anomaly has the following variation with
respect to the background metric $\ro$
\be
\delta_{\ro}\GAM_{\!\mbox{\tiny Chiral}}[\mu,\mub;\ro]\ =\ \sm{k}{12 \pi}\lp
\prt\zer{\nabla}\mu + \prtb\zer{\ovl{\nabla}}\mub
-\prt\prtb\ln\!\ro \rp \delta\ln\!\ro\ .
\lbl{varo}
\ee

\indent

In concluding this section, one may say that the construction of a
Polyakov action heavily relies on the use of complex 
coordinates pertaining to the complex structure
defined by $\mu$ together with the integrating factor as well 
introduced by R.~Stora in \cite{Sto87,Sto88}.
The logarithm of the integrating factor turns out to be deeply related to the
renormalization problem, namely it plays a crucial role in the search
of a resummation for the Feynman perturbative series. As conjectured in 
\cite{LS89,Laz90}, it has a universal charactere as
defining the renormalized UV-behaviour at
coincident points of the Feynman propagator $\ln|z-w|^2$ for free
fields in the plane.

\sect{On the energy-momentum tensor in 2-d conformal field theory}

\indent
 
It is well known that in a Lagrangian approach the energy-momentum 
is related to the Schwarzian derivative $\{Z,z\}$
\cite{GN89,AS89,Hab90} and quite recently recalled in \cite{AT97}.
However, in this Lagrangian view point, $\mu$ and $\mub$ are sources 
for the two components $\Theta$ (respectively $\ovl{\Theta}$)  
of the energy momentum-tensor and yield the holomorphic factorization 
property of the correlation functions of the latter \cite{BPZ84,BB87,Bec88}. 
Geometrically one ought to expect that $Theta$
is a true (2,0)-conformal field, namely a quadratic differential.
In order to compute $\delta_{\mu}\GAM_{\!\mbox{\tiny
Chiral}}[\mu,\mub;\ro]$ it will be technically much more convenient
to rewrite the effective action \rf{WZP} as
\bea
&&\GAM_{\!\mbox{\tiny Chiral}}[\mu,\mub;\ro] \ =\ 
\sm{k}{12\pi}\int_{\Sigma} \mes{z}{z}\! \blp 
\!\shalf \lp \ln\!\ro-\ln(\lambda \ovl{\lambda}) \rp \!
\prt\prtb\lp \ln(\lambda \ovl{\lambda})-\ln\!\ro \rp 
-\prtb\ln\lambda\prt\ln\ovl{\lambda} \nn\\[2mm]
&&\hs{-15} +\ \mu(\Ro-\{Z,z\}) + 
\prtb\ln\lambda\, (\prt\ln\!\ro - \prt\ln\lambda) 
+ \mub(\zer{\ovl{R}\,}-\{\bZ,\bz\}) 
+ \prt\ln\ovl{\lambda}\, (\prtb\ln\!\ro - \prtb\ln\ovl{\lambda}) \brp,
\lbl{WZP'}
\ena
after the use of the identity due to equation \rf{lam}, 
$\prtb\ln\lambda = \cD\mu$,
\bea
\lp \ln(\lambda \ovl{\lambda})-\ln\!\ro \rp \!
\prt\prtb\ln(\lambda \ovl{\lambda}) &=&
\prt \left[ \cD\mu \lp 
\ln (\lambda \ovl{\lambda}) - \ln\!\ro \rp \right]
+ \prtb \left[ \ovl{\cD}\mub \lp \ln(\lambda \ovl{\lambda})-\ln\!\ro
\rp \right] \nn\\[2mm]
&&\hs{-40}+\ (\prt\ln\lambda-\prt\ln\ro) \cD\mu + 
(\prtb\ln\ovl{\lambda}-\prtb\ln\ro) \ovl{\cD}\mub
- 2 \cD\mu\ovl{\cD}\mub.
\ena
Note that the first term in \rf{WZP'} is a Liouville-like piece
for the Liouville (scalar) field \rf{Liouphio} and connects the two
chiral sectors through the background metric. 
Then, it is straightforward to show that (modulo a well-defined total
differential)
\be
\delta_\mu\GAM_{\!\mbox{\tiny Chiral}}[\mu,\mub;\ro]\ =\
\sm{k}{12\pi}(\Ro-\{Z,z\})\delta\mu\ ,
\lbl{e-m-0}
\ee
where from eq.\rf{lam} $(\prtb-\mu\prt)\delta\ln\!\lambda = \cD\delta\mu$.
The energy-momentum tensor \rf{e-m-0} turns out to be holomorphically
factorized as a non-local current in the complex structure $\mu$. 
Geometrically, as expected, thanks to the introduction of a background metric,
it is a smooth quadratic differential with respect
to the background complex coordinates $\zbz$ as a difference of two
(smooth) projective connections. Accordingly in the
background metric, one has the following anomalous
conformal Ward identities for the diffeomorphism symmetry
\be
(\prtb-\mu\prt-2\prt\mu)\frac{\GAM_{\!\mbox{\tiny
Chiral}}[\mu,\mub;\ro]}{\delta\mu}
= \sm{-k}{12\pi}\lp \prt^3 \mu 
- (\prtb - \mu \prt - 2\prt \mu ) \Ro \rp \ ,
\lbl{WI0}
\ee
since $(\prtb - \mu \prt - 2\prt \mu )\{Z,z\} = \prt^3\mu$, and where
the r.h.s. of \rf{WI0} is the integrand of the global anomaly \rf{a-0}.

\sect{Background Liouville field versus background holomorphic
projective connection}

\indent

The above Polyakov formula \rf{WZP'} is supposed to be a
resummation of the perturbative expansion in the correlation functions.
Remark that the vacuum expectation value (VEV) of $\Theta$ depends on the
background metric by quantum action principle, one has
\be
<\Theta>_{\ro} = \left. 
\frac{\GAM_{\!\mbox{\tiny Chiral}}[\mu,\mub;\ro]}{\delta\mu}
\right|_{\mu=\mub=0} = \sm{k}{12\pi}\Ro,
\lbl{VEV0}
\ee
where the expression expression on the r.h.s. 
$\Ro\ =\prt^2\ln\!\ro - \shalf (\prt\ln\!\ro)^2$, has alredy been
proposed by D.~Friedan in \cite{Fri84}.
The property of holomorphy, $\prtb<\Theta>_{\ro}=0$,
is achieved if and only if the background metric $\ro$ fulfills the
following globally defined Liouville equation
\be
\prt\prtb\ln\!\ro -\, a\!\ro\ =\ 0\ ,
\lbl{Liouv}
\ee
with real constant $a$, that is the background metric is of 
constant scalar curvature. Recall
that $\ro$ belongs to the conformal class of the flat metric. On an
arbitrary compact Riemann surface, one can convince oneself that the
Liouville equation \rf{Liouv} turns out to be equivalent to the
holomorphy of the curvature $\Ro$, namely $\prtb\Ro\ =0$ \cite{Sto92}.

Therefore, in the case where the background metric is governed by the
Liouville equation \rf{Liouv}, the chirally split diffeomorphism
anomaly \rf{a-0} writes
\be
\anom(C,\mu;\ro)\ =\ \sm{k}{12 \pi} \int_{\Sigma} \mes{z}{z} 
C \lp \prt^3 + 2\Ro\!\prt + \prt\Ro \rp\mu 
\lbl{a-00}
\ee
and the energy-momentum tensor has a holomorphic VEV as required.

However, if it is not the case, that is $\Ro$ is only smooth, it is
customary to introduce a background holomorphic projective connection
${\cal R}$ by adding a further globally defined chirally split
local counterterm \cite{KLT90}
\be
\GAM_{\!\mbox{\tiny II}}(\mu,\mub;\ro,{\cal R},\ovl{\cal R})\ =\ 
\sm{k}{12 \pi} \int_{\Sigma} \mes{z}{z} 
\mu ({\cal R} - \Ro) \ + \ \mbox{c.c.} \ ,
\ee
to the chrially split functional. More precisely, in this latter situation 
formula \rf{WZP'} is replaced by 
$$\GAM_{\!\mbox{\tiny Chiral}}[\mu,\mub;\ro,{\cal R},\ovl{\cal R}] \equiv
\GAM_{\!\mbox{\tiny Chiral}}[\mu,\mub;\ro] + 
\GAM_{\!\mbox{\tiny II}}(\mu,\mub;\ro,{\cal R},\ovl{\cal R})$$ 
with the replacement of $\Ro$ by $\cR$,
\bea
&&\GAM_{\!\mbox{\tiny Chiral}}[\ro,{\cal R},\ovl{\cal R},\mu,\mub] \ =\ 
\sm{k}{12\pi}\int_{\Sigma} \mes{z}{z}\! \blp 
\!\shalf \lp \ln\!\ro-\ln(\lambda \ovl{\lambda}) \rp \!
\prt\prtb\lp \ln(\lambda \ovl{\lambda})-\ln\!\ro \rp
-\prtb\ln\lambda\prt\ln\ovl{\lambda} \nn\\[2mm]
&&\hs{-15} +\ \mu({\cal R}-\{Z,z\}) + 
\prtb\ln\lambda\, (\prt\ln\!\ro - \prt\ln\lambda) 
+ \mub(\ovl{{\cal R}}-\{\bZ,\bz\}) 
+ \prt\ln\ovl{\lambda}\, (\prtb\ln\!\ro - \prtb\ln\ovl{\lambda}) \brp,
\lbl{WZPproj}
\ena
while a direct computation yields
\be
s\,\GAM_{\!\mbox{\tiny Chiral}}[\mu,\mub;\ro,{\cal R},\ovl{\cal R}]\ =\ 
\anom(C,\mu;{\cal R}) \ 
+\ \overline{\anom(C,\mu;{\cal R})} \ ,
\lbl{a-plit}
\ee
with the standard chirally split diffeomorphism anomaly 
independent of $\ro$ recalled in \rf{anom},
and for the variations, on the one hand,
\be
\delta_{\ro}\GAM_{\!\mbox{\tiny Chiral}}[\mu,\mub;\ro,{\cal R},\ovl{\cal R}]
\ =\ \sm{-k}{12 \pi}\prt\prtb\ln\!\ro\,\delta\ln\!\ro
\ =\ \sm{k}{48\pi}R(\go)\,\delta\!\ro\,
\lbl{ro0}
\ee
is proportional to the scalar curvature 
${\dps R(\go)=\frac{-4}{\ro} \prt\prtb\ln\!\ro}$ 
of the backgroung metric $\go$, and on the other hand,
\be
\delta_\mu\GAM_{\!\mbox{\tiny Chiral}}[\mu,\mub;\ro,{\cal R},\ovl{\cal R}]\
=\ \sm{k}{12\pi}({\cal R}-\{Z,z\})\delta\mu\ .
\lbl{e-m}
\ee
Hence the corresponding anomalous conformal Ward identities now write
\be
(\prtb-\mu\prt-2\prt\mu)\frac{\GAM_{\!\mbox{\tiny
Chiral}}[\mu,\mub;\ro,{\cal R},\ovl{\cal R}]}{\delta\mu}
\ =\ \sm{-k}{12\pi}\lp \prt^3 
+ 2{\cal R}\prt + \prt{\cal R}\rp\mu  \ ,
\lbl{WI}
\ee
while the VEV for the energy-momentum tensor now is
\be
<\Theta>_{{\cal R}}\ =\ \left. 
\frac{\GAM_{\!\mbox{\tiny Chiral}}[\mu,\mub;\ro,{\cal R},\ovl{\cal R}]}{\delta\mu}
\right|_{\mu=\mub=0} =
\sm{k}{12\pi}{\cal R},
\lbl{VEV}
\ee
which does no longer depend on the background metric.
Both the latter formulae can respectively be deduced from \rf{WI0} and
\rf{VEV0} by the direct substitution $\Ro\lra{\cal R}$.

However, note that the independence on the background of $\GAM_{\!\mbox{\tiny
Chiral}}$ is no longer achieved. This is certainly the signature of the 
ill-treatment of the zero mode problem.

The background feature defined by the smooth metric
$\ro$ has been partially shifted into the favour of a background holomorphic
projective connection ${\cal R}$. Recall either the latter survives at the
level of the anomaly problem or the former occurs through the combination
$\Ro$, see \rf{WI} or \rf{WI0} respectively. 
In any case, the background(s) $\ro$
(and ${\cal R}$) is (are) strongly related to the details of the
model, besides the central charge $k$, and encodes the global effects
for instance in the VEV for the stress-energy tensor, say.

\sect{Conclusion and outlook}

\indent

This very naive approach yields a well-defined holomorphically
factorized formula valid on any higher genus compact Riemann surface
without boundary where a background metric links the two chiral sectors.
This formula reproduces the well-known Polyakov action in the plane.
However, the starting point with the usual Liouville
action was very wrong. Recall that in \cite{KLS91a,KLS91b}
the independence in the Weyl factor is crucial and is achieved 
to the benefit of quantities depending on the complex analytic geometry only. 
Here the dependence on the background is the
signature of the ill-treatment of the zero mode problem and especially
the difficulty in defining a propagator on a curved space.

Performing directly in $\ZBZ$ a holomorphic change of coordinates,
$F(Z)$, formula \rf{WZP} receives non-trivial boundary terms depending
on $\ln F'(Z)$, the latter vanish if $F$ is affine, \ie the Riemann
surface has null Euler characteristic (torus). For higher genus, the
problem of finding a Polyakov action remains open and call for further
investigation, especially in the use of the genuine Wess-Zumino
action. In order to define the perturbative series, one should start
with the difference of two such functionals integrating the Weyl anomaly
\cbe
\int_{\Sigma}\mes{Z}{Z} \blp\shalf \Phi\paZ\Pab\Phi + \Phi\ln\!\rZ\brp
- \int_{\Sigma}\mes{z}{z} \blp\shalf \Phi\prt\prtb\Phi + \Phi\ln\!\rho\brp,
\cee
and try to express the scalar field $\Phi$ as a non-local functional
of the Beltrami coefficient $\mu$ through its equation of motion. In
doing this difference, the problem of zero modes ought to be 
resolved \cite{Dow94}.
Also, links with the WZNW approach as treated in \cite{Yos90} must be
considered.

\vskip 1.cm \noindent
{\bf Acknowledgements :} The author wishes to thank the organizers
of this 1-st worshop in Mathematical Physics held in Montpellier
as well as some financial support from the french GDR682 for the
opportunity all they gave him to confront some impressions about
the gains and losses of understanding in Liouville theory out of which
the present work has been developed.


\begin{thebibliography}{10}
\bibliographystyle{unsrt}

\bibitem{Pol87}
A.M. Polyakov.
\newblock Quantum gravity in two dimensions.
\newblock {\em \MPL{A2}}, pages 893--898, (1987).

\bibitem{Yos89}
K.~Yoshida.
\newblock Effective action for quantum gravity in two dimensions.
\newblock {\em \MPL{A4}}, pages 71--81, (1989).

\bibitem{BBG89}
L.~Baulieu, M.~Bellon, and R.~Grimm.
\newblock Left-right asymmetric conformal anomalies.
\newblock {\em \PLB{228}}, pages 325--331, (1989).

\bibitem{Laz90}
S.~Lazzarini.
\newblock {\em Sur les mod\`eles conformes lagrangiens bidimensionnels}.
\newblock PhD thesis, Universit\'e de Savoie and LAPP-TH, April 1990.
\newblock unpublished, in french.

\bibitem{Gri90}
R.~Grimm.
\newblock Left-right decomposition of two-dimensional superspace geometry and
  its {BRS} structure.
\newblock {\em Ann. Phys. {\bf 200}, 1}, pages 49--100, (1990).

\bibitem{Hab90}
Z.~Haba.
\newblock Generating functional for the energy-momentum tensor in
  two-dimensional conformal field theory.
\newblock {\em \PR{D 41}}, pages 724--726, (1990).

\bibitem{Yos90}
K.~Yoshida.
\newblock Effective action for induced gravity on a two-dimensional manifold.
\newblock {\em \IJMP{A5}}, pages 4559--4578, (1990).

\bibitem{Zuc91}
R.~Zucchini.
\newblock A {P}ol\-ya\-kov action on {R}ie\-mann sur\-fa\-ces.
\newblock {\em \PLB{260}}, pages 296--298, (1991).

\bibitem{Zuc93a}
R.~Zucchini.
\newblock A {P}ol\-ya\-kov action on {R}ie\-mann sur\-fa\-ces {(II)}.
\newblock {\em Commun. Math. Phys. {\bf 152}}, pages 269--298, (1993).

\bibitem{AT97}
E.~Aldrovandi and L.A. Takhtajan.
\newblock Generating functional in {CFT} and effective action for
  two-dimensional quantum gravity on higher genus {R}ie\-mann sur\-fa\-ces.
\newblock {\em Commun. Math. Phys. {\bf 188}}, pages 29--67, (1997).
\newblock {\tt hep-th/9606163}.

\bibitem{Pol81}
A.M. Polyakov.
\newblock {Q}uantum geometry of bosonic strings.
\newblock {\em \PLB{103}}, pages 207--210, (1981).

\bibitem{KLS91a}
M.~Knecht, S.~Lazzarini, and R.~Stora.
\newblock On holomorphic factorization for free conformal fields.
\newblock {\em Phys. Lett. {\bf B 262}}, pages 25--31, (1991).

\bibitem{KLS91b}
M.~Knecht, S.~Lazzarini, and R.~Stora.
\newblock On holomorphic factorization for free conformal fields {II}.
\newblock {\em Phys. Lett. {\bf B 273}}, pages 63--66, (1991).

\bibitem{BGS88}
J.-M. Bismut, H.~Gillet, and C.~Soul\'e.
\newblock Analytic torsion and holomorphic determinant bundles, {I}, {II} {\&}
  {III}.
\newblock {\em \CMP{115}}, pages 49, 79 {\&} 301, (1988).

\bibitem{Gie93b}
F.~Gieres.
\newblock On the holomorphic factorization for superconformal fields.
\newblock {\em \LMP{28}}, pages 43--48, (1993).
\newblock {\tt hep-th/9209078}.

\bibitem{BK86a}
A.A. Belavin and V.G. Knizhnik.
\newblock Algebraic geometry and the geometry of quantum strings.
\newblock {\em \PLB{168}}, pages 201--206, (1986).

\bibitem{BK86b}
A.A. Belavin and V.G. Knizhnik.
\newblock Complex geometry and the theory of quantum strings.
\newblock {\em Sov. Phys. JETP {\bf 64}}, pages 214--228, (1986).

\bibitem{BBS86}
L.~Baulieu, C.~Becchi, and R.~Stora.
\newblock ``{O}n the covariant quantization of the free bosonic string''.
\newblock {\em \PLB{180}}, pages 55--60, (1986).

\bibitem{Bec88}
C.M. Becchi.
\newblock ``{O}n the covariant quantization of the free string: the conformal
  structure''.
\newblock {\em Nucl. Phys. {\bf B304}}, page 513, (1988).

\bibitem{Leh87}
O.~Lehto.
\newblock {\em Univalent functions and {T}eichm{\"u}ller spaces}.
\newblock Graduate Texts in Mathematics. Springer-Verlag, Berlin, 1987.

\bibitem{Sto88que}
R.~Stora.
\newblock Differential algebras in field theory.
\newblock Unpublished Lectures given at the 27-th Session of S\'eminaire de
  Math\'ematiques Sup\'erieures de l'Universit\'e de Montr\'eal on ``Methods in
  field and string theories'', (Qu\'e) Canada, June 6-24, 1988.

\bibitem{GN89}
J.~Grundberg and R.~Nakayama.
\newblock On 2d (super) gravity in the light-cone gauge.
\newblock {\em \MPL{A4}}, page~55, (1989).

\bibitem{BB87}
L.~Baulieu and M.~Bellon.
\newblock ``{B}eltrami parametrization in string theory''.
\newblock {\em Phys. Lett. {\bf B196}}, page 142, (1987).

\bibitem{Sto87}
R.~Stora.
\newblock Alg{\`e}bres diff{\'e}rentielles en th{\'e}orie des champs.
\newblock {\em Ann. Inst. Fourier {\bf 37}, 4}, pages 235--245, (1987).

\bibitem{KLT90}
M.~Knecht, S.~Lazzarini, and F.~Thuillier.
\newblock Shifting the {W}eyl anomaly to the chirally split diffemorphism
  anomaly in two dimensions.
\newblock {\em Phys. Lett. {\bf B251}}, pages 279--283, (1990).

\bibitem{Laz97}
S.~Lazzarini.
\newblock Flat complex vector bundles, the {B}eltrami differential and
  {W}-algebras.
\newblock {\em \LMP{41}}, pages 207--225, (1997).
\newblock {\tt hep-th/9802083}.

\bibitem{BC-RR83}
L.~Bonora, P.~Cotta-Ramusino, and C.~Reina.
\newblock Conformal anomaly and cohomology.
\newblock {\em \PLB{126}}, pages 305--308, (1983).

\bibitem{LS89}
S.~Lazzarini and R.~Stora.
\newblock Ward identities for {L}agrangian conformal models.
\newblock In L.~Lusanna, editor, {\em Knots, Topology and Quantum Field
  Theory}, 13-th John Hopkins workshop. World Scientific, 1989.

\bibitem{Gun66}
R.C. Gunning.
\newblock {\em Lectures on {R}iemann Surfaces}.
\newblock Princeton Mathematical Notes. Princeton Univ. Press, Princeton, New
  Jersey, 1966.

\bibitem{WZ71}
J.~Wess and B.~Zumino.
\newblock Consequences of anomalous {W}ard identities.
\newblock {\em \PLB{37}}, pages 95--97, (1971).

\bibitem{SSW89}
M.N. Sanielevici, G.W. Semenoff, and Y.S. Wu.
\newblock Path integral analysis of chiral bosonization.
\newblock {\em \NPB{312}}, pages 197--226, (1989).

\bibitem{VV89}
E.~Verlinde and H.~Verlinde.
\newblock {\em Conformal field theory and geometric quantization.}
\newblock PUPT-89-1149 or IASSNS-HEP-89/58, October 1989.
\newblock Based on lectures given at the Trieste Spring School, April 1989, and
  at the Ecole Normale Sup{\'e}rieure in Paris, June 1989. Published in Trieste
  Superstrings 1989:422-449.

\bibitem{Ver90}
H.~Verlinde.
\newblock Conformal field theory, two-dimensional quantum gravity and
  quantization of {T}eichm{\"u}ller space.
\newblock {\em \NPB{337}}, pages 652--680, (1990).

\bibitem{Sto88}
R.~Stora.
\newblock The role of locality in string theory.
\newblock In {\em Non perturbative quantum field theory}, NATO ASI Ser.B,
  Vol.{\bf 185}. Carg{\`e}se Summer School Institute, July 1987, Plenum Press,
  1988.

\bibitem{AS89}
A.~Alekseev and S.~Shatashvili.
\newblock Path integral quantization of the coadjoint orbits of the {V}irasoro
  group and 2-d gravity.
\newblock {\em \NPB{323}}, pages 719--733, (1989).

\bibitem{BPZ84}
A.M. Belavin, A.M. Polyakov, and A.B. Zamolodchikov.
\newblock Infinite conformal symmetry in two-dimensional quantum field theory.
\newblock {\em \NPB{241}}, pages 333--380, (1984).

\bibitem{Fri84}
D.~Friedan.
\newblock Introduction to {P}olyakov's string theory.
\newblock In {\em Recent Advances in Field Theory and Statistical Mechanics},
  Les Houches Session XXXIX, 1982, pages 839--867, Amsterdam, 1984. Elsevier
  Science Publishers.

\bibitem{Sto92}
R.~Stora.
\newblock Private communication.

\bibitem{Dow94}
J.S. Dowker.
\newblock A note on {P}olyakov's non-local form of the effective action.
\newblock {\em Class. Quant. Grav. {\bf 11}}, pages L7--L9, (1994).

\end{thebibliography}
\end{document}